\documentclass[twocolumn,showpacs,pre]{revtex4}
\usepackage{dcolumn}
\usepackage{graphicx}
\usepackage{graphicx,amssymb,amsmath,revsymb}
\usepackage{natbib}
\usepackage{tikz}
\usepackage{color}

\newcommand{\x}{\tilde{t}}
\newcommand{\y}{\tilde{v}}

\begin{document}

\title{Non-equilibrium structure and dynamics in a microscopic model of thin film active gels}

\author{D. A. Head$^{1}$, W. J. Briels$^{2}$ and Gerhard Gompper$^{3}$}

\affiliation{$^{1}$School of Computing, Leeds University, Leeds LS2 9JT, UK.}
\affiliation{$^{2}$Computational Biophysics, University of Twente, 7500 AE Enschede, The Netherlands,}
\affiliation{$^{3}$Theoretical Soft Matter and Biophysics, Institute of Complex Systems and Institute for Advanced Simulation, Forschungszentrum J\"ulich, 52425 J\"ulich, Germany.}

\date{\today}

\begin{abstract}
In the presence of ATP, molecular motors generate active force dipoles that drive suspensions of protein filaments far from thermodynamic equilibrium, leading to exotic dynamics and pattern formation.
Microscopic modelling can help to quantify the relationship between 
individual motors plus filaments to organisation and dynamics on molecular 
and supra-molecular length scales.
Here we present results of extensive numerical simulations of active gels where the motors and filaments are confined between two infinite parallel plates. Thermal 
fluctuations and excluded-volume interactions between filaments are included. 
A systematic variation of rates for motor motion, attachment and detachment, 
including a differential detachment rate from filament ends, reveals a range of non-equilibrium behaviour. Strong motor binding produces structured filament aggregates that we refer to as asters, bundles or layers, whose stability depends on motor speed and differential end-detachment. The gross features of the dependence of the observed structures on the motor rate and the filament concentration can be captured by a simple one-filament model.
Loosely bound aggregates exhibit super-diffusive mass transport, where filament translocation scales with lag time with non-unique exponents that depend on motor kinetics.
An empirical data collapse of filament speed as a function of motor speed and end-detachment is found, suggesting a dimensional reduction of the relevant parameter space. We conclude by discussing the perspectives of microscopic modelling in the field of active gels.
\end{abstract}


\maketitle

%
%
\section{Introduction}

Mixtures of protein filaments and molecular motors form an established class of active media, in which spontaneous internal processes drive the system from thermodynamic equilibrium~\cite{Ramaswamy2010}. Protein filaments and molecular motors represent the dynamic intracellular scaffolding known as the cytoskeleton that performs a range of tasks crucial to organism viability~\cite{HowardBook,BoalBook,BrayBook,AlbertsBook}. That similar phenomena to those observed {\em in vivo} can be reproduced in systems lacking genetic control~\cite{Heald1996,Daga2006,Nurse2006} suggests that some form of self-organisation has been exploited by 
natural selection to robustly produce beneficial phenotypes. Identifying and elucidating the principles of self-organisation relevant to these active gels will therefore increase our understanding of the  processes that sustain life, and leave us better equipped to counteract defects when they arise. 

Mesoscopic theoretical models are uniquely placed to investigate 
such phenomena, as they permit hypothesis testing unconstrained by experimental limitations, and full, non-invasive data extraction.  A range of theories based on a continuum description of the 
local director field of filament orientation, which assume that variations 
are slow on the length scale of single filaments (the so-called 
``hydrodynamic" limit),
have now been devised~\cite{Marchetti2013,Aranson2003,Aranson2005,Liverpool2006,Cates2008,Giomi2010,Tjhung2012,Sankararaman2009}, including those based on nematodynamic~\cite{Voituriez2005,Voituriez2006,Basu2008,Elgeti2011} and Smoluchowski~\cite{Liverpool2003,Ahmadi2005,Zeibert2005,Ruehle2008} approaches, predicting a range of self-organised pattern formation as control parameters are varied. For example, asters, nematic phases, density instabilities, and vortices have been predicted and qualitatively observed.

A recognised deficiency of such ``hydrodynamic" models is their dependence on 
phenomenological parameters that cannot be easily related to 
molecular mechanisms. Microscopic models can bridge these length scales, but most devised to date neglect steric hinderance between filaments, from which nematic elasticity derives and without which many of the predicted states cannot be realised~\cite{Surrey2001,Ziebert2008,Pinot2009,Loughlin2010,Wang2011,Kohler2011,Saintillan2012}. The reason for this omission may be due to the specific application considered, but may also be simply pragmatic, as incorporating excluded-volume interactions in numerical simulation 
is notoriously expensive. Coupled with the high aspect ratio of filaments, making it difficult to achieve linear system sizes much larger than the filament length at reasonable densities,  means numerical simulation of active gels is a formidable challenge. Analytical coarse graining is therefore desirable, but has so far only been performed for rigid, adamant motors which do not induce relative filament rotation~\cite{Liverpool2005} (here adamant refers to a motor's insensitivity to loading, which has been argued to make spontaneous flow impossible~\cite{Wang2011}).

The potential benefits to be made from microscopic modelling motivates its continued pursuit, even if results are limited for now to relatively small systems. For strictly two dimensional (2D) systems, anomalous diffusion and 
large-wavelength density fluctuations were observed~\cite{Head2011a} as in models of active media~\cite{Tu1998,Chate2008,Golestanian2009,Ramaswamy2003}, but structural self-organisation was inhibited by the steric hinderance, resulting in disordered structures unlike {\em in vitro} experiments~\cite{Surrey2001}. 3D systems confined between parallel plates reduce steric hinderance by allowing a degree of filament overlap without excessively increasing the numerical burden. With additional lateral confinement in a ring-like corral (representing either the cell membrane or the effect of other filaments around), spindle-like configurations and rotating vortices were observed in delineated regions of parameters space comprised of motor speed and density~\cite{Head2011b}.

Here we consider quasi-2D active gels confined between parallel planes with periodic boundaries in the lateral directions, in order to describe active systems in thin films without lateral confinement.
Our aim is to elucidate and quantify structure and dynamics on molecular 
and supra-molecular length scales, and how they result from the 
various microscopic parameters.
We systematically vary the end-detachment rate to control the dwell time of motors at filament ends, which is sometimes incorporated into models lacking excluded volume~\cite{Ziebert2008,Pinot2009} where it has been argued to be necessary to reproduce vortices~\cite{Surrey2001}. Both the mean filament speed and the exponents describing anomalous diffusion are sensitive to end-detachment as detailed in Sec.~\ref{s:resDynamics}. This is argued to be due to motor motility being limited by loading, and the load in turn dominated by static motors dwelling at filament ends. For high motor densities, many-filament clusters form that can be classified into asters, layers and bundles as described in Sec.~\ref{s:resStatics}. The layered state, strikingly reminiscent of microtubule structures that self-organise from {\em Xenopus} cytosol~\cite{Mitchison2013}, is only clearly defined when end-detachment is enhanced, confirming the importance of end-dwelling in guiding the motor-driven self-organisation.
 It is also dynamically stable in the presence of thermal noise, similar to 
active smectics and other striped non-equilibrium steady 
states~\cite{Adhyapak2013}.
The observed trends are reproduced in a 
simple, effective one-filament model that supports this 
interpretation. Finally, in Sec.~\ref{s:discussion} we discuss how close we are to achieving our goal of reaching experimental length and time scales 
{\em in silico}, and suggest possible means to close the gap.

%
%
\section{Model}
\label{s:methods}

The model is referred to as microscopic as the shortest length represented is no larger than the dimensions of individual motors or filaments. The model is explained below first in terms of the components, then the method used to integrate the system in the specified geometry is detailed.

\subsection{Motors and filaments}

Filaments are modelled as linear arrays of $M=30$ monomers with centers spaced by a distance~$b$ as shown in Fig.~\ref{f:schematic}(a). The filaments are 
polar and have $[-]$- and $[+]$-ends that define the direction of motor motion. 
The unit vector from $[-]$ to $[+]$ is denoted~$\hat{\bf p}$. Steric hinderance between filaments is incorporated as repulsive forces acting between non-bonded monomers, here taken to be a Lennard-Jones potential parameterised by an energy $\varepsilon=5k_{\rm B}T$ and a length scale $\sigma=b$, with a cut-off at the potential minimum $r=2^{1/6}\sigma$~\cite{FrenkelSmit,AllenTildesly}. The filament length is $L=Mb$, and mapping this to the protein fiber in question allows $b$ to be estimated, {\em e.g.}, $b\approx100nm$ for a 3$\mu$m protein filament.

Bipolar motor clusters (hereafter simply called motors) are only explicitly represented when attached to filaments. Soluble motors are instead implicitly incorporated into the fixed attachment rate $k_{\rm A}$ (this simplification, which can be relaxed~\cite{Surrey2001}, corresponds to an infinite reservoir of soluble motors). Motors only attach to pairs of monomers of different filaments with centers within a specified distance, here taken to be the same as the interaction range $2^{1/6}b$. Once attached, the motor is represented as a two-headed spring, with each head located at the center of the attached monomer. The spring constant is $k_{\rm B}T/b^{2}$ and the natural spring length is $2^{1/6}b$ so that they attach in an (almost) unstressed state.

Motor heads move by one or more monomers at a time in the direction of the filament's $[+]$-end, as shown in Fig.~\ref{f:schematic}(b). Since the distance of order $b$ per step will typically be much larger than the step size of real motor proteins~\cite{HowardBook}, this should be regarded as the integration of a series of smaller movements. Motor loading exponentially retards motion according to the change $\Delta E$ in motor elastic energy that would be induced by the move,
\begin{equation}
\left\{
\begin{array}{l@{\quad:\quad}c}
k_{\rm M}e^{-\Delta E/k_{\rm B}T} & \Delta E\geq0\:, \\
k_{\rm M} & \Delta E<0\:.
\end{array}
\right.
\label{e:move}
\end{equation}
The form of Eq.~(\ref{e:move}) suppresses moves that would increase the motor spring energy too much, acting as a stall force. $k_{\rm M}$ corresponds to the unloaded motor rate, which has been tabulated for real proteins~\cite{HowardBook}.
Moves of more than one monomer are allowed but are exponentially rare due to their typically high~$\Delta E>0$. Each motor head detaches at a rate $k_{\rm D}$, in which event the entire motor is removed from the system. 
Motor heads residing at a filament's $[+]$-end detach at a rate $k_{\rm E}$ which may differ from $k_{\rm D}$; see Fig.~\ref{f:schematic}(c). 
Finally, motors do not move if by doing so they would exceed 
a maximum head-to-head separation of~$5b$; however, if overstretching 
(head-to-head separation larger than ~$5b$) is induced by the relative motion 
of the filaments, then overstretched motors are removed from the filaments.

\subsection{Iteration}

The filament positions and orientations are updated as per the Brownian dynamics of rigid rods~\cite{DoiEdwards}. For each time step~${\rm d}t$, all forces (motor-mediated plus excluded volume) acting on each filament are summed to give the total force ${\bf F}$ and torque ${\bf W}$. These are then converted to a change in the filament center-of-mass vector ${\bf x}^{\rm COM}$ as
\begin{eqnarray}
\delta {\bf x}^{\rm COM}
&=&
\frac{1}{\gamma^{\parallel}} \left[\xi_{1} \sqrt{2\gamma^{\parallel}kT{\rm d}t} \:\:+  {\bf F}\cdot\hat{\bf p}\:\:\,{\rm d}t\right] \hat{\bf p}
\nonumber\\
&+&
\frac{1}{\gamma^{\perp}}\left[\xi_{2} \sqrt{2\gamma^{\perp}kT{\rm d}t}+  {\bf F}\cdot\hat{\bf n}_{1}\,{\rm d}t\right] \hat{\bf n}_{1}
\nonumber\\
&+&
\frac{1}{\gamma^{\perp}}\left[\xi_{3} \sqrt{2\gamma^{\perp}kT{\rm d}t}+  {\bf F}\cdot\hat{\bf n}_{2}\,{\rm d}t\right] \hat{\bf n}_{2}\:,
\label{e:modelTrans}
\end{eqnarray}
where the $\xi_{i}$ are uncorrelated random variables drawn from a unit Gaussian distribution, and the unit vectors $\hat{\bf n}_{1}$ and $\hat{\bf n}_{1}$ are chosen at each time step such that  $(\hat{\bf p},\hat{\bf n}_{1},\hat{\bf n}_{2})$ form an orthonormal basis. The damping coefficients are related to the drag coefficient $\gamma$ of an individual monomer by $\gamma^{\parallel}=M\gamma$, $\gamma^{\perp}=2\gamma^{\parallel}$. The filament is then rotated about its new centre-of-mass to give a new orientation unit vector $\hat{\bf p}^{\rm new}=(\hat{\bf p}+\delta{\bf p})/|\hat{\bf p}+\delta{\bf p}|$, where
\begin{eqnarray}
\delta{\bf p}
=
\frac{1}{\gamma_{M}}
\Big{[}
{\bf W}\times\hat{\bf p}\,{\rm d}t
&+&
\xi_{4}\sqrt{2kT\gamma_{M}{\rm d}t}\,\hat{\bf n}_{1}
\nonumber\\
&+&
\xi_{5}\sqrt{2kT\gamma_{M}{\rm d}t}\,\hat{\bf n}_{2}
\Big{]}
\label{e:modelRot}
\end{eqnarray}
where $\gamma_{M}=\frac{1}{12}M(M^{2}-1)b^{2}\cdot2\gamma$ plays the role of the moment of inertia in this overdamped system. The bead positions are then updated according to the new ${\bf x}^{\rm COM}$ and~$\hat{\bf p}$. The use of rigid rods deviates from previous work where the filaments were flexible, which required a smaller $\delta t$ for numerical stability~\cite{Head2011a,Head2011b}.

\subsection{Geometry and numerical procedure}

\begin{figure}[htpb]
\centerline{\includegraphics[width=8.5cm]{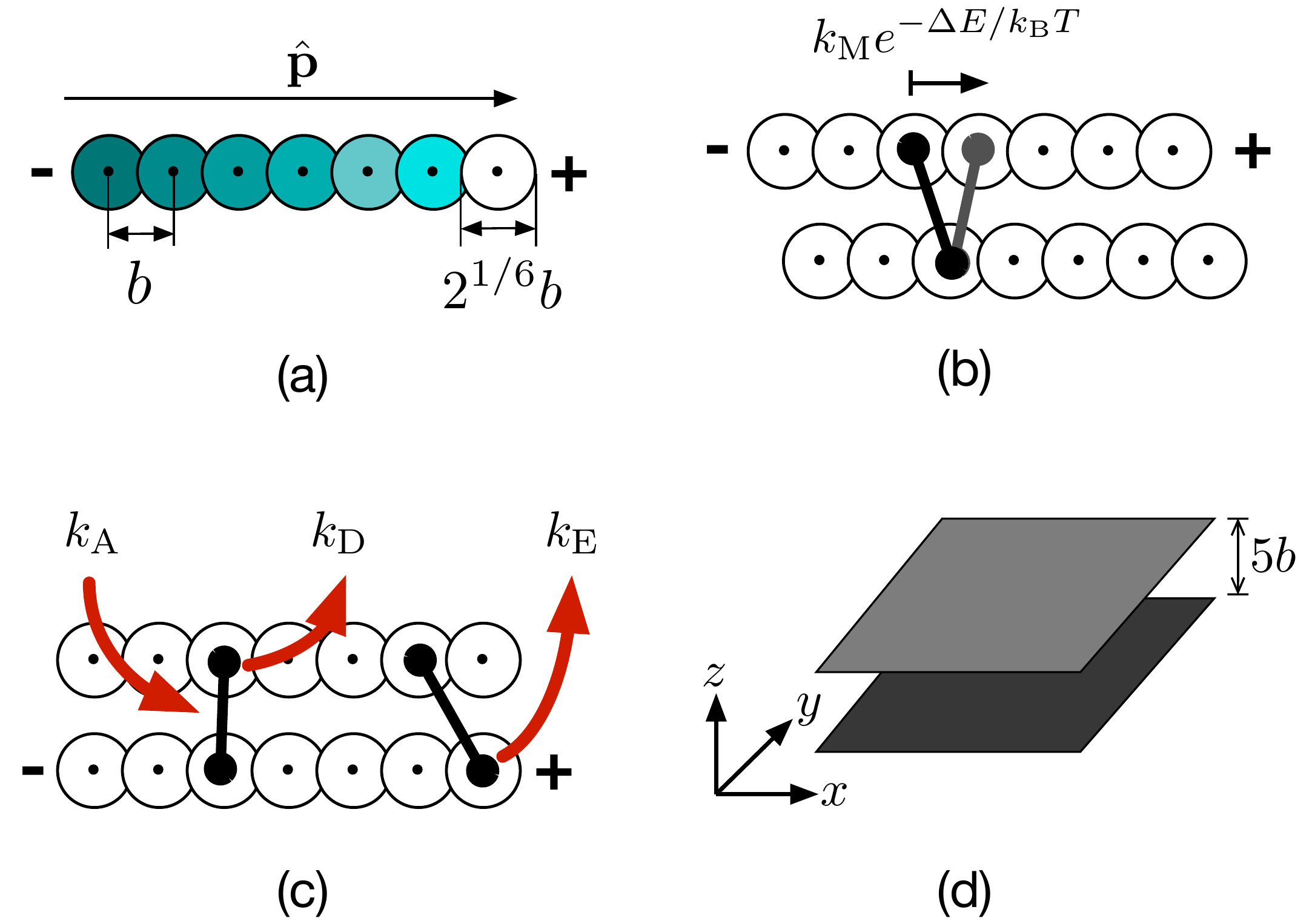}}
\caption{(a)~Filaments are linear monomer arrays with centers $b$ apart, with a polarity vector $\hat{\bf p}$ directed from $[-]$ to $[+]$. Each monomer has an excluded-volume interaction of range $2^{1/6}b$ to non-bonded monomers. 
(b)~Each motor head moves at a rate $k_{\rm M}e^{-\Delta E/k_{\rm B}T}$ if the corresponding increase in spring energy $\Delta E\geq 0$; for $\Delta E<0$, the rate is simply~$k_{\rm M}$. 
(c)~Motors attach at a rate $k_{\rm A}$ when the monomers are within a prescribed distance. Each head detaches at a rate $k_{\rm D}$, leading to removal of the motor. If the head is at the $[+]$-end, this rate becomes~$k_{\rm E}$. (d)~The system is narrowly confined in the $z$-direction, with periodic boundaries for  $x$ and~$y$.
}
\label{f:schematic}
\end{figure}

The system has dimensions $(X,Y,Z)$ with $X=Y=125b\approx4L$ and $Z=5b=L/6$ as shown in Fig.~\ref{f:schematic}(d). The system is periodic in the $x$ and $y$-directions, but there are repulsive walls along the planes $z=0$ and $z=Z$ with the same potential and parameters as the excluded-volume interactions. 
As $Z\ll L$, these walls restrict filament orientations to lie approximately in the $x$-$y$ plane while still permitting overlap. The density of the system is given in terms of the volume fraction $\phi=N v_{\rm f}/XYZ$ for $N$ filaments of volume $v_{\rm f}$ each, where $v_{\rm f}$ is the volume of a cylinder of diameter $2^{1/6}\sigma$ with hemispherical end-caps.

Convergence with time was checked by ensuring a sample of measured quantities (nematic order parameter, motor density and mean squared displacements) were independent of time. Densities above $\phi\approx0.2$, or motor speeds below $k_{\rm M}\approx k_{\rm D}$, did not reach stationarity within the attainable simulation times of around $10^{2}k_{\rm D}^{-1}$ and were avoided. Motor speeds above $k_{\rm M}\approx10^{3}k_{\rm D}$ placed a finite fraction of motors close to their maximum extension, resulting in a significant rate of motor breakage through overextension under relative filament motion. These speeds were also avoided to reduce the number of mechanisms under consideration.

%
%
\section{Results}
\label{s:results}

Snapshots representative of the parameter space sampled are presented in Fig.~\ref{f:snapshots}. Movies are provided in the supplementary information~\cite{SuppInf}. Filament configurations can be broadly identified as belonging to one of two groups: 
(i) weakly bound states of small, transient clusters, or 
(ii) strongly bound states with spatially-extended structure formation. The former class displays a range of exotic dynamics and is the subject of Sec.~\ref{s:resDynamics}. The motor-driven structure formation for strongly bound states is detailed in Sec.~\ref{s:resStatics}, and is supported by analysis of a 
simple, effective one-filament model that highlights the 
controlling role of $k_{\rm E}$ in selecting between aster and layer states.

All results are presented in dimensionless form by scaling lengths by the filament length $L=Mb$, and times or rates by either the detachment rate $k_{\rm D}$ or the time $\tau_{\rm L}=M/k_{\rm M}$ for an unloaded motor to traverse a filament. The relationship to the equivalent experimental scales is discussed in Sec.~\ref{s:discussion}.

\begin{figure}[htbp]
\centerline{\includegraphics[width=8.5cm]{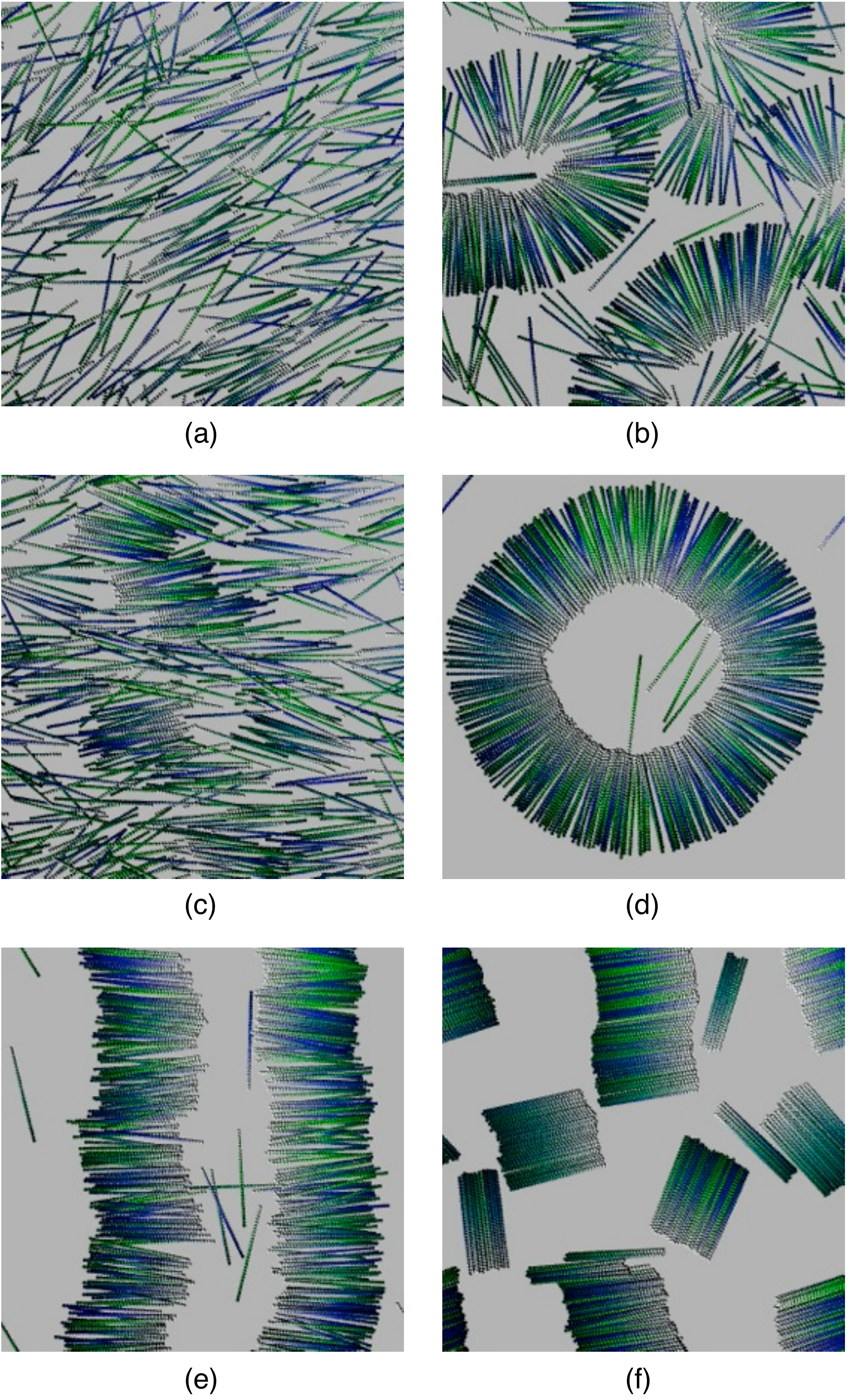}}
\caption{Snapshots representative of regions of parameter space for weakly-bound states with $k_{\rm A}=20k_{\rm D}$ in (a)--(c), and strongly bound states with $k_{\rm A}=40k_{\rm D}$ in (d)--(f). The other parameters are
(a)~$k_{\rm E}/k_{\rm D}=10$, $k_{\rm M}=10^{2}k_{\rm D}$ and $\phi=0.1$,
(b)~$k_{\rm E}/k_{\rm D}=1$, $k_{\rm M}=10^{2}k_{\rm D}$ and $\phi=0.15$,
(c)~$k_{\rm E}/k_{\rm D}=5$, $k_{\rm M}=10^{2}k_{\rm D}$ and $\phi=0.15$.
(d)~$k_{\rm E}/k_{\rm D}=1$, $k_{\rm M}=10^{2}k_{\rm D}$ and $\phi=0.15$,
(e)~$k_{\rm E}/k_{\rm D}=5$, $k_{\rm M}=10^{2}k_{\rm D}$ and $\phi=0.15$ and
(f)~ $k_{\rm E}/k_{\rm D}=1$, $k_{\rm M}=k_{\rm D}$ and $\phi=0.2$.
Light (dark) shades correspond to filament $[+]$ ($[-]$)-ends. Motors are not shown for reasons of clarity, but are provided in the matching figures in the supplementary information, along with movies for the same parameters~\cite{SuppInf}.
}
\label{f:snapshots}
\end{figure}


%
%
\subsection{Dynamics of weakly-bound states}
\label{s:resDynamics}

A basic dynamic quantity is the mean filament translational speed 
$v^{\rm RMS}\equiv\sqrt{\langle v^{2}\rangle}$ averaged over particle 
trajectories in steady state~\cite{Saintillan2012}. However, instantaneous velocities are not 
well defined for overdamped dynamics with thermal noise, as employed here 
[see Eqs.~(\ref{e:modelTrans}) and (\ref{e:modelRot})]. It is therefore 
necessary to estimate the velocity over a finite time interval~$t>0$, 
but this raises further difficulties since filament motion is not ballistic 
in the regimes of interest, {\em i.e.} the displacement vector 
$\Delta {\bf x}(t)\equiv{\bf x}(t_{0}+t) - {\bf x}(t_{0})$ of a filament 
center ${\bf x}$ is not linear in~$t$, making it difficult to define a unique velocity. Instead, 
we first consider a nominal speed defined over a fixed time interval, 
$v^{\rm RMS}\equiv\Delta r(t^{\rm RMS})/t^{\rm RMS}$ with 
$\Delta r\equiv|\Delta {\bf x}|$ and $t^{\rm RMS}=(4k_{\rm D})^{-1}$, 
as a measure of net motility, and consider trends with respect to variations 
in $k_{\rm M}$ and $k_{\rm E}$. Varying $t^{\rm RMS}$ alters the values 
of $v^{\rm RMS}$ but not these trends. The full spectrum of displacements 
with varying lag times is then considered in more detail.

\begin{figure}
\centerline{\includegraphics[width=8.5cm]{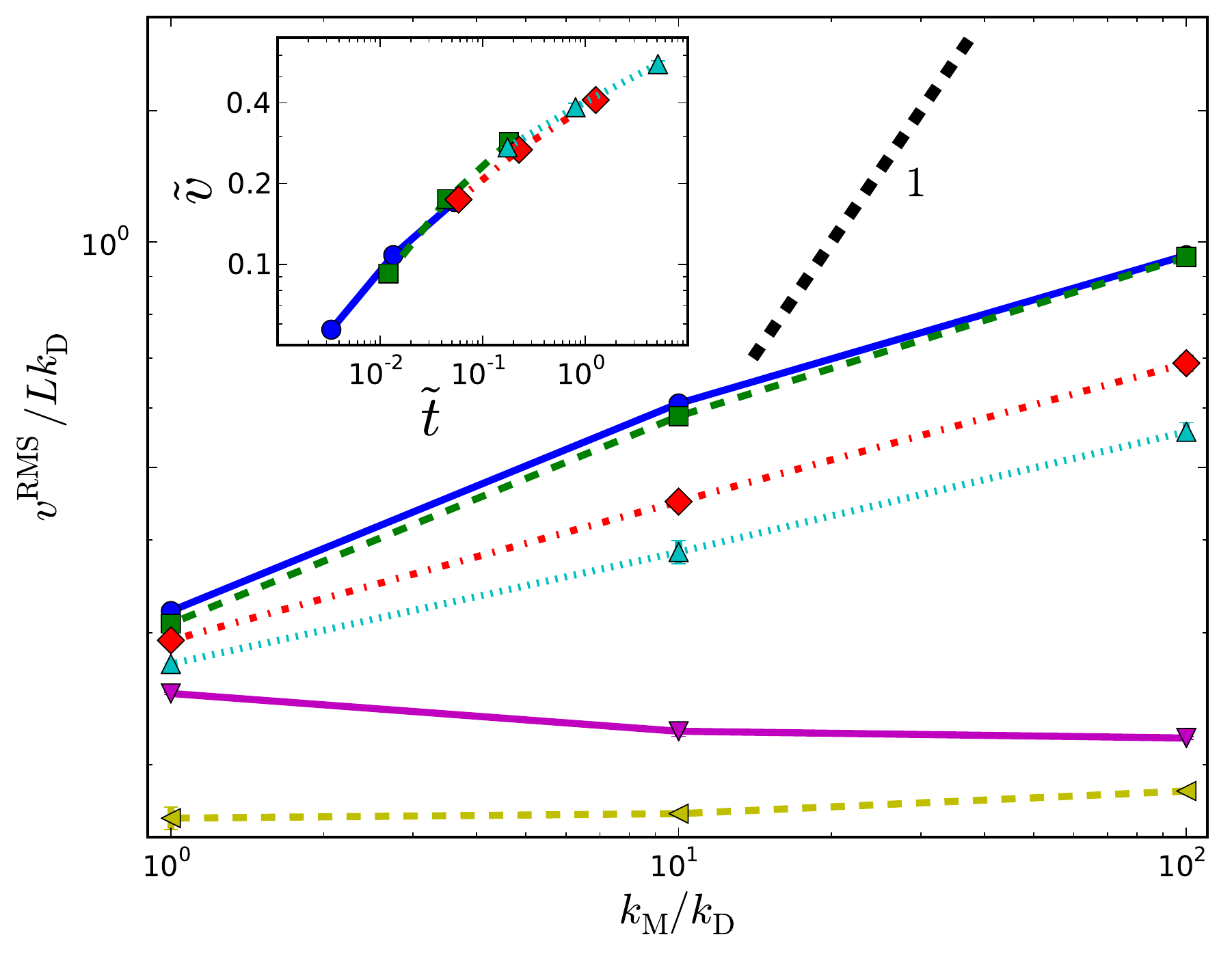}}
\caption{Filament speed $v^{\rm RMS}$ versus unloaded motor speed $k_{\rm M}$
in a double-logarithmic representation, for (from bottom to top) $k_{\rm E}/k_{\rm D}$=0.2, 0.5, 1, 2, 5 and 10 respectively. $k_{\rm A}=20k_{\rm D}$, $\phi=0.15$ and the thick dashed line has a slope of 1. The inset shows the scaled velocity $\y=(k_{\rm D}/k_{\rm E})^{3/4}(v^{\rm RMS}/Lk_{\rm D})$ against the scaled dwell time $\x=(t_{\rm occ}^{[+]}/t_{\rm occ})(k_{\rm D}/k_{\rm E})$ for the $k_{\rm E}\geq k_{\rm D}$ data points only.
}
\label{f:velMags}
\end{figure}

Fig.~\ref{f:velMags} shows $v^{\rm RMS}$ versus $k_{\rm M}$ for a range of $k_{\rm E}$ from $k_{\rm E}=0.2k_{\rm D}$ to $k_{\rm E}=10k_{\rm D}$.
For $k_{\rm E}<k_{\rm D}$, the system forms a strongly-bound aster state similar to Fig.~\ref{f:snapshots}(d), and correspondingly low values of~$v^{\rm RMS}$. Such states are the focus of Sec.~\ref{s:resStatics} and will not be pursued further here. For $k_{\rm E}\geq k_{\rm D}$, states more closely resemble Figs.~\ref{f:snapshots}(a-c) and
$v^{\rm RMS}$ monotonically increases 
with $k_{\rm M}$ but at a slower rate than the naive expectation 
$v^{\rm RMS}\propto k_{\rm M}$, which would arise from a filament
being pulled with constant motor stepping rate $k_{\rm M}$ across other filaments.
Sub-linear scaling of speed with activity (controlled via ATP concentration) 
has also been inferred from experiments~\cite{Sanchez2012,Thampi2013}.
Possible origins of this sub-linear behavior are that for larger
$k_{\rm M}$ motors more often reach their stall force, or are experiencing more
frequent force-induced detachments from the filament. Furthermore,
the observation from Fig.~\ref{f:velMags} that $v^{\rm RMS}$ increases with~$k_{\rm E}$ suggests end-dwelling motors act to suppress filament motion. To test this hypothesis, let $t^{[+]}_{\rm occ}$ denote the mean dwell time of motor heads at $[+]$-ends, and $t_{\rm occ}$ the occupancy time at any other point along the filament ({\em i.e.}, before the head detaches or moves). All of the $v^{\rm RMS}$ can be collapsed onto a single-valued function of $t^{[+]}_{\rm occ}/t_{\rm occ}$ after rescaling both axes by powers of $k_{\rm E}/k_{\rm D}$. As demonstrated in Fig.~\ref{f:velMags} (inset), good collapse arises when employing the scaling variables $\x=(t^{[+]}_{\rm occ}/t_{\rm occ})(k_{\rm D}/k_{\rm E})$ and $\y=(k_{\rm D}/k_{\rm E})^{3/4}(v^{\rm RMS}/Lk_{\rm D})$, {\em i.e.}, $\y=g(\x)$ with scaling function $g$. 
That $v^{\rm RMS}$ is a function of $k_{\rm E}/k_{\rm D}$ and the relative dwell time at $[+]$-ends, confirms system activity is strongly influenced by end-dwelling. The origin of the scaling exponents for $\x$ and $\y$ are not yet evident.

\begin{figure}
\centerline{\includegraphics[width=8.5cm]{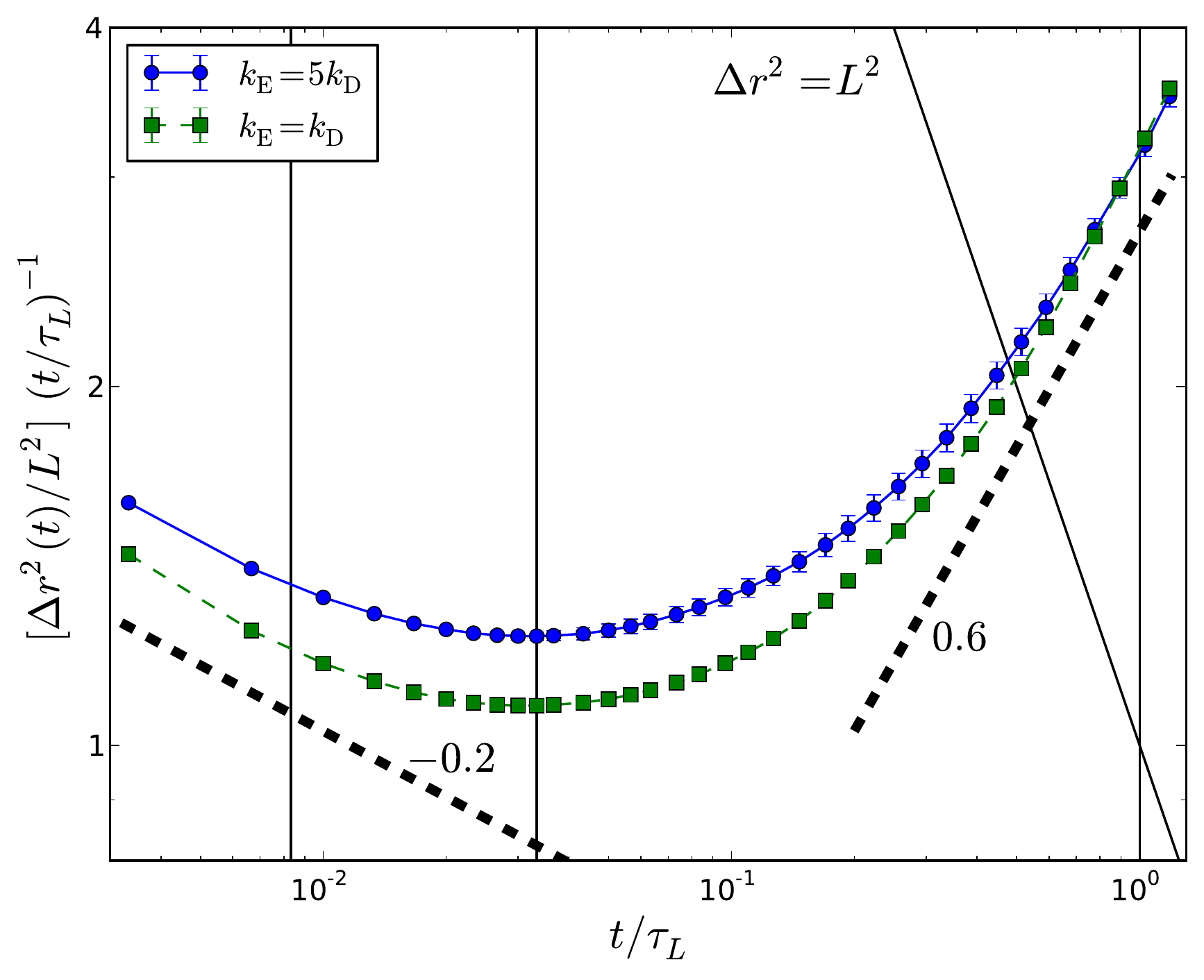}}
\caption{
Mean squared displacements $\Delta r^{2}(t)$ versus lag time $t$ plotted such that normal 
diffusion $\Delta r^{2}\propto t$ corresponds to a horizontal line. 
$k_{\rm A}=10k_{\rm D}$, $k_{\rm M}=k_{\rm D}$, $\phi=0.15$, and the $k_{\rm E}$ are given 
in the legend. The thin diagonal line corresponds to displacements equal to the filament length, {\em i.e.} $\Delta r^{2}=L^{2}$. The thick dashed lines, which have slopes $-0.2$ and $0.6$ on these axes, correspond to sub-diffusion $\Delta r^{2}\propto t^{0.8}$ and super-diffusion $\Delta r^{2}\propto t^{1.6}$ respectively. The leftmost vertical line corresponds to $t=t^{\rm RMS}$, {\em i.e.} the time interval used to calculate the $v^{\rm RMS}$ in Fig.~\ref{f:velMags}. The middle and rightmost vertical lines correspond to $t=k_{\rm M}^{-1}$ and $t=Mk_{\rm M}^{-1}\equiv \tau_{L}$ respectively.
}
\label{f:msd_kA10}
\end{figure}

\begin{figure}
\centerline{\includegraphics[width=8.5cm]{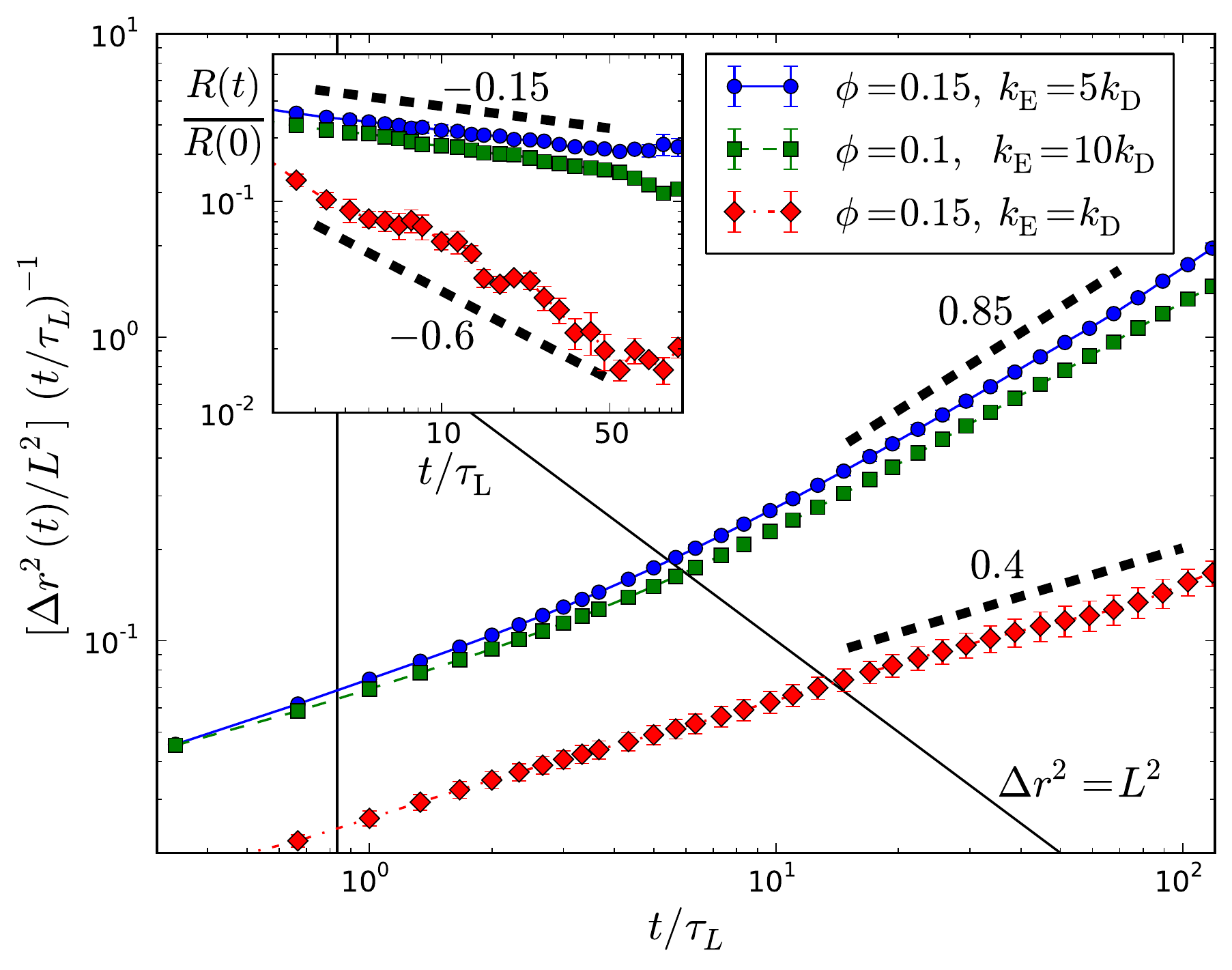}}
\caption{
Mean squared displacements $\Delta r^{2}(t)$ versus lag time $t$ plotted in the same manner 
as in Fig.~\ref{f:msd_kA10}. $k_{\rm A}=20k_{\rm D}$, $k_{\rm M}=10^{2}k_{\rm D}$, and 
$\phi$ and $k_{\rm E}$ are given in the legend. The thin diagonal line corresponds to displacements equal to the filament length, $\Delta r^{2}=L^{2}$. The thin vertical line corresponds to $t=t^{\rm RMS}$.
(Inset)~The velocity autocorrelation function $R(t)=\langle{\bf v}(0)\cdot{\bf v}(t)\rangle$ for the same runs. For both plots, the thick dashed lines have the given slopes.
}
\label{f:msd_vv}
\end{figure}

Extending this analysis to self-diffusion reinforces the important role of end-dwelling. Active media often exhibit super-diffusion with mean-squared displacements $\Delta r^{2}$ that vary super-linearly with time, $\Delta r^{2}\propto t^{a}$ with $1<a\leq2$, as observed in intracellular transport~\cite{Bursac2005,Zhou2009,Bruno2009}, {\em in vitro} experiments~\cite{Kohler2011,Kohler2012,Sanchez2012} and models of self-propelled particles~\cite{Tu1998,Chate2008,Golestanian2009}. Conversely, $0<a<1$ is referred to as sub-diffusion. Both forms of anomalous diffusion have been measured in our model, as shown in Fig.~\ref{f:msd_kA10} which gives $\Delta r^{2}(t)$ for weakly bound systems $k_{\rm A}=10k_{\rm D}$. Sub-diffusion with $a\approx0.8$ is observed over short times $t$ when the motors are acting as passive crosslinkers, generating viscoelasticity of the aggregate structures that retards filament motion~\cite{Mason2000}. For larger~$t$, when motor motion becomes relevant, a crossover to super-diffusion with $a\approx1.6$ is clearly seen. This super-diffusive regime becomes more dominant with a higher density of motors, as shown in Fig.~\ref{f:msd_vv} for the higher $k_{\rm A}=20k_{\rm D}$. Further increasing $k_{\rm A}$ generates strongly-bound structures such as Figs.~\ref{f:snapshots}(d)--(f), which remain sub-diffusive for the largest simulation times achieved.

\begin{figure}
\centerline{\includegraphics[width=8.5cm]{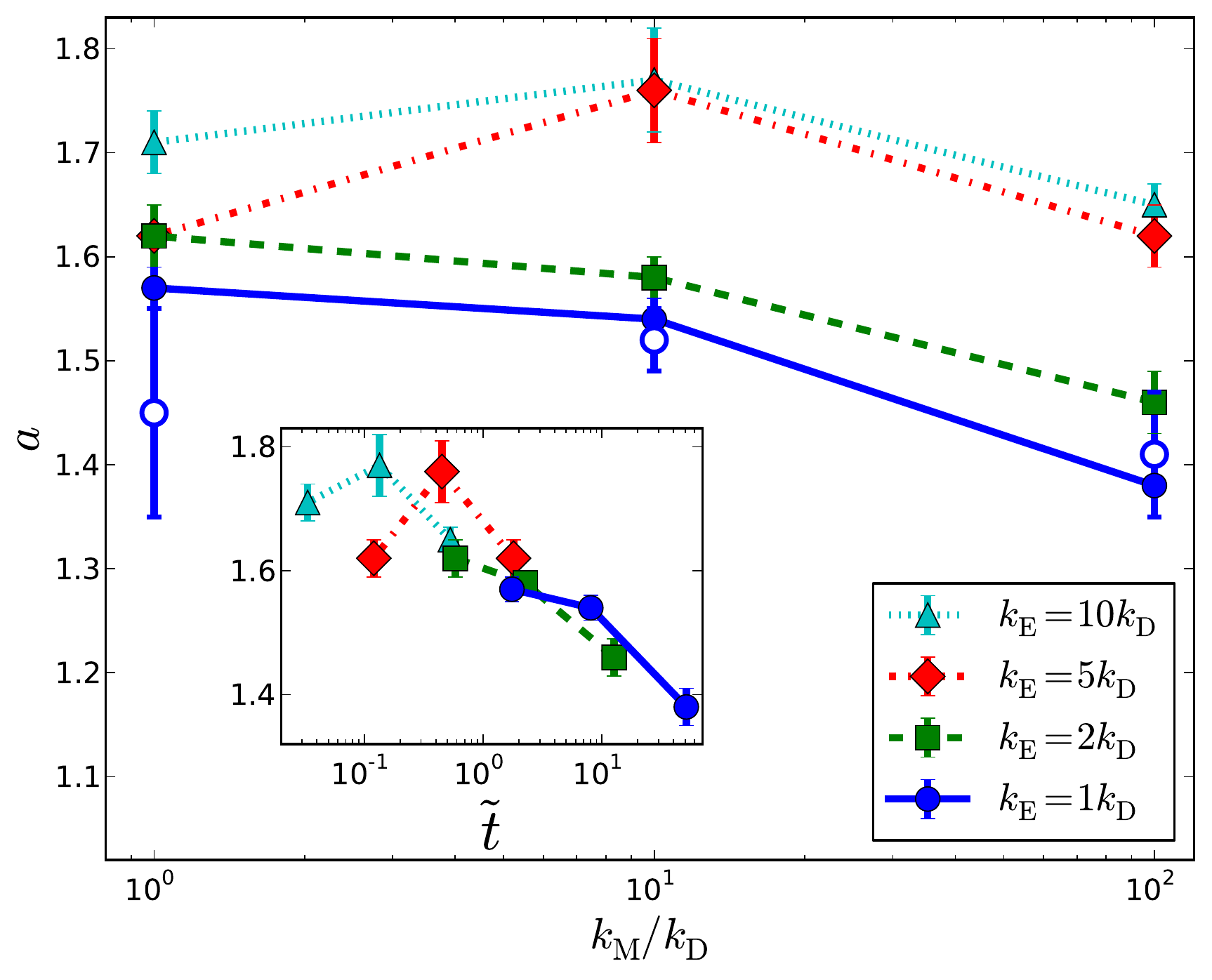}}
\caption{Effective MSD exponent $a$ for the mean-squared displacements (filled symbols) versus $k_{\rm M}$, for $\phi=0.15$, $k_{\rm A}=20k_{\rm D}$ and the $k_{\rm E}$ given in the legend. The open symbols show $a$ as measured from the decay of velocity autocorrelations $R(t)\sim t^{a-2}$ for $k_{\rm E}=k_{\rm D}$ (other $k_{\rm E}$ not shown for clarity but give similar agreement). (Inset)~Data plotted against the same $\x$ as in Fig.~\ref{f:velMags}, demonstrating partial collapse.}
\label{f:msd_exps}
\end{figure}

Independent evaluation of $a>1$ is possible from the velocity autocorrelation function $R(t)\equiv\langle {\bf v}(0)\cdot{\bf v}(t)\rangle$, which in steady state obeys~\cite{Taylor1922,Majda1999}
\begin{equation}
\langle\Delta r^{2}(t)\rangle
=
2
\int_{0}^{t}{\rm d}s\,(t-s)R(s)\:,
\label{e:vv}
\end{equation}
from which it immediately follows that $1<a\leq2$ corresponds to $R(t)\sim t^{a-2}$. $R(t)$ is plotted in Fig.~\ref{f:msd_vv} (inset) and is consistent with this prediction.
The exponent~$a$, as determined from fitting $\Delta r^{2}$ at the same length $\Delta r^{2}=L^{2}$, for a range of $k_{\rm E}$ and $k_{\rm M}$ is shown in Fig.~\ref{f:msd_exps}, and is seen to cover a similar range to that measured for intra-cellular traffic~\cite{Bursac2005,Bruno2009}. The variation with $k_{\rm M}$ is non-monotonic; however, $a$ monotonically increases with the end-detachment rate $k_{\rm E}$, and for high $k_{\rm E}$ approaches $a=2$ as observed in reconstituted active gels~\cite{Kohler2011,Kohler2012,Sanchez2012}. This observation suggests end-dwelling is again playing a key role, and plotting the exponent against the same scaling variable $\x$ as above collapses the data as shown in the figure inset. Although here the collapse is only partial, the significant clustering compared to the unscaled data demonstrates the importance of end-dwelling.

\begin{figure}
\centerline{\includegraphics[width=6.5cm]{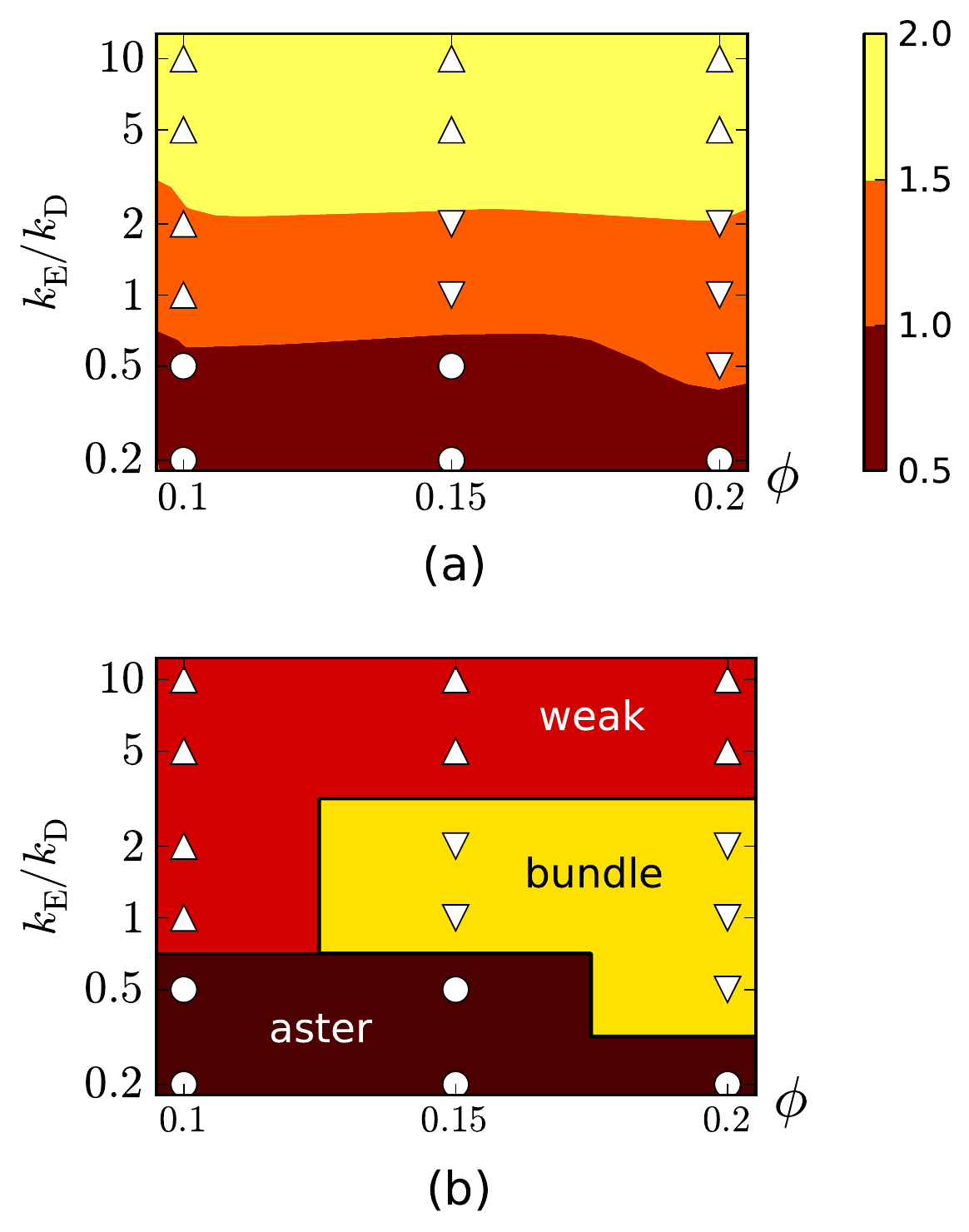}}		
\caption{
(a)~Effective MSD exponent and (b) state as a function of $\phi$ and $k_{\rm E}/k_{\rm D}$ for $k_{\rm A}=20k_{\rm D}$ and $k_{\rm M}=10^{2}k_{\rm D}$. Symbols denote actual data points and contours are linearly interpolated. The calibration bar for (a) denotes the value of the MSD exponent. The state was determined using the procedure described in Sec.~\ref{s:resStatics}.
}
\label{f:stateMSDlowkA}
\end{figure}

The variation of the effective MSD exponent~$a$ with $k_{\rm E}$ and $\phi$ is presented in Fig.~\ref{f:stateMSDlowkA}, where we also plot the state of these same data points using the procedure to be described in Sec.~\ref{s:resStatics}. High filament density and low $k_{\rm E}$ give rise to persistent, localised clusters such as those evident in Fig.~\ref{f:snapshots}(b) and~(c), which are termed bundles. Such states, although super-diffusive with~$a>1$, have a much lower exponent than the nematic states that arise for high $k_{\rm E}$ or low~$\phi$, which are referred to as weak binding in the figure, and resemble Fig.~\ref{f:snapshots}(a).
Asters predominately form for $k_{\rm E}<k_{\rm D}$ for this $k_{\rm A}$ 
and $k_{\rm M}$, with correspondingly subdiffusive dynamics with $a<1$ as 
seen in the figure.

\begin{figure}
\centerline{\includegraphics[width=8.5cm]{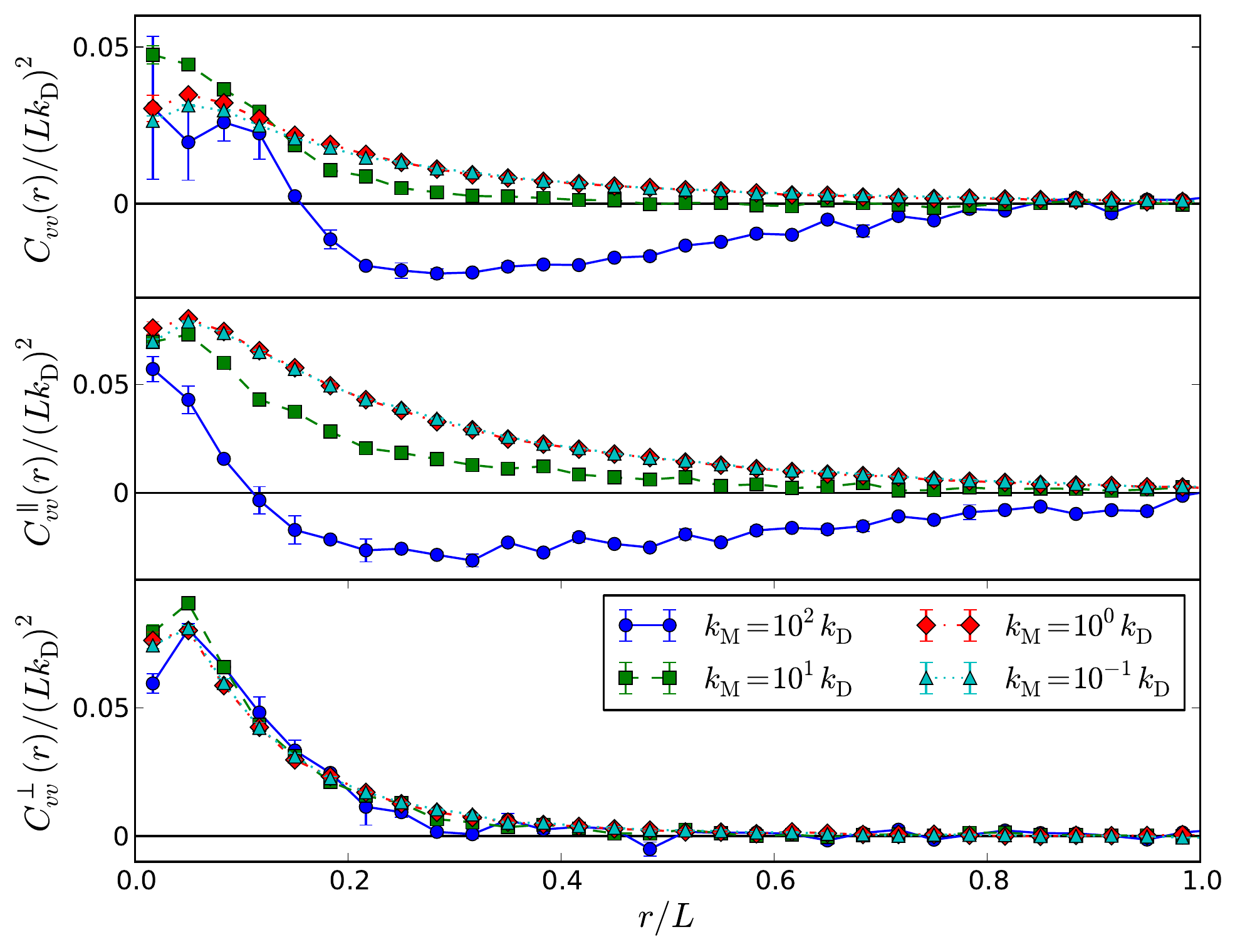}}
\caption{Spatial velocity correlations without projection $C_{vv}(r)$, and projected parallel and perpendicular to the filament polarity vector, $C_{vv}^{\parallel}(r)$ and $C_{vv}^{\perp}(r)$ respectively, for the $k_{\rm M}$ given in the legend in the lower panel, $k_{\rm A}=20k_{\rm D}$, $k_{\rm E}=5k_{\rm D}$ and $\phi=0.15$.
}
\label{f:spatVel}
\end{figure}

Spatial correlations in velocity reveal instantaneous modes of relative 
filament motion, and has been used to quantify the effect of mutations on 
cytoplasmic streaming {\em in vivo}~\cite{Ganguly2012}, and of ATP concentration on active flow {\em in vitro}~\cite{Sanchez2012} and in ``hydrodynamic" models~\cite{Thampi2013}. The two-point correlation function $C_{vv}(r)=\langle{\bf v}(0)\cdot{\bf v}({\bf r})\rangle$ provides information purely as a function of the distance $r=|{\bf x}^{\beta}-{\bf x}^{\alpha}|$ separating filament centres ${\bf x}^{\alpha}$ and ${\bf x}^{\beta}$. Additional insight can be gained by projecting the separation vector parallel and perpendicular to the filament polarity $\hat{\bf p}^{\alpha}$, {\em i.e.},
\begin{equation}
C^{\parallel}_{vv}(r)
=
\frac
{\sum_{\alpha,\beta}\,{\bf v}^{\alpha}\cdot{\bf v}^{\beta}\,\delta(r-|{\bf x}^{\alpha}-{\bf x}^{\beta}|)\cos^{2}\theta}
{\sum_{\alpha,\beta}\delta(r-|{\bf x}^{\alpha}-{\bf x}^{\beta}|)\cos^{2}\theta},
\label{e:Cvv_par}
\end{equation}
where $\cos\theta=\hat{\bf p}^{\alpha}\cdot({\bf x}^{\beta}-{\bf x}^{\alpha})/r$. The corresponding expression for $C^{\perp}_{vv}(r)$ is given by replacing $\cos^{2}\theta$ by $\sin^{2}\theta$. (Using $\hat{\bf p}^{\beta}$ to calculate $\theta$ gives the same result due to the symmetry of Eq.~(\ref{e:Cvv_par})). The variation of $C_{vv}(r)$, $C^{\parallel}_{vv}(r)$ and $C^{\perp}_{vv}(r)$ with $k_{\rm M}$ is plotted in Fig.~\ref{f:spatVel}, and exhibits qualitatively different behavior for the two projections:~$C^{\perp}_{vv}$ is always positive, while $C^{\parallel}_{vv}$ exhibits a negative region for fast motors. This trend remains true for all $1\leq k_{\rm E}/k_{\rm D}\leq10$ considered, with a broader anti-correlated region for increasing~$k_{\rm E}$ when filament motion is less inhibited. Throughout this range the filament polarity vectors are aligned in parallel, as evident in the corresponding polarity correlation functions described in Sec.~\ref{s:resStatics}. Inspection of Eq.~(\ref{e:Cvv_par}) then reveals that $C^{\parallel}_{vv}<0$ corresponds to contrary motion of overlapping filaments. Cytoplasmic streaming in {\em Drosophila} egg cells exhibited anti-correlations over lengths of approximately $18\mu$m, comparable to the microtubule length~\cite{Ganguly2012}, and therefore on longer lengths than observed here. In addition, little or no variation in correlation length with motor speed was observed either in the {\em Drosophila} system, or in reconstituted {\em in vitro} networks  and a ``hydrodynamic" model~\cite{Sanchez2012,Thampi2013}, unlike the variation apparent in Fig.~\ref{f:spatVel}. The cause of this deviation is not clear, but may simply be due to the smaller systems studied here not permitting active swirls to fully develop.
It may also be due to the lack of hydrodynamic interactions in our model, 
which has been shown to give long-range velocity correlations in a 
microscopic model~\cite{Saintillan2012}.

%
%

%
%
\subsection{Structure formation for strong binding}
\label{s:resStatics}

\begin{figure}
\centerline{\includegraphics[width=8.5cm]{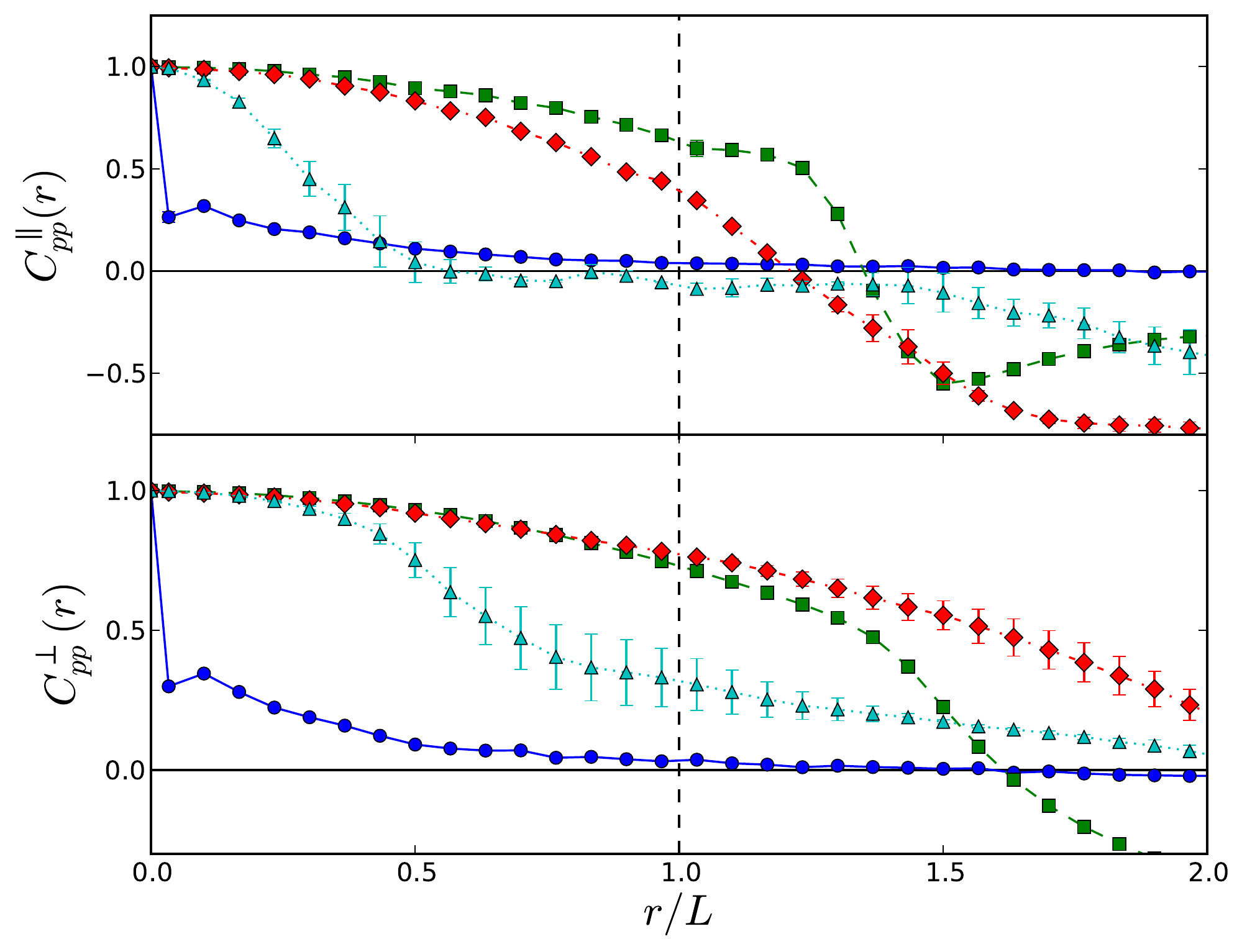}}
\caption{The polarity correlation function projected parallel $C^{\parallel}_{pp}(r)$ (top) and perpendicular $C^{\perp}_{pp}(r)$ (bottom) to the filament axis.
Symbols refer to the same parameters in Fig.~\ref{f:snapshots}:
Circles to Fig.~\ref{f:snapshots}(a) (weakly bound),
squares to Fig.~\ref{f:snapshots}(d) (aster), 
diamonds to Fig.~\ref{f:snapshots}(e) (layer) and
triangles to Fig.~\ref{f:snapshots}(f) (bundle).
}
\label{f:polarCorrns}
\end{figure}

Increasing the motor density, {\em e.g.}, by raising the attachment rate~$k_{\rm A}$, produces extended clusters consisting of many filaments. Three distinct configurations were observed in this strongly-bound regime for the parameter space sampled, namely asters, layers and bundles as demonstrated in Fig.~\ref{f:snapshots}(d), (e) and~(f) respectively. Signatures of the structural organisation are apparent in the spatial correlations in filament polarity~$\hat{\bf p}$, quantified by projecting relative displacements parallel and perpendicular to the filament axis analogously to the velocity correlations~(\ref{e:Cvv_par}). Plots of both 
$C_{pp}^{\parallel}(r)$ and $C_{pp}^{\perp}(r)$ are given in 
Fig.~\ref{f:polarCorrns} for examples of each of the three states mentioned, and also for a weakly bound state by way of comparison. Projecting the correlations in this manner, rather than using a single 
averaged quantity~\cite{Saintillan2012,Sanchez2012,Thampi2013}, provides 
additional information which can be used to extract the structure formation.

\begin{figure}
\centerline{\includegraphics[width=6cm]{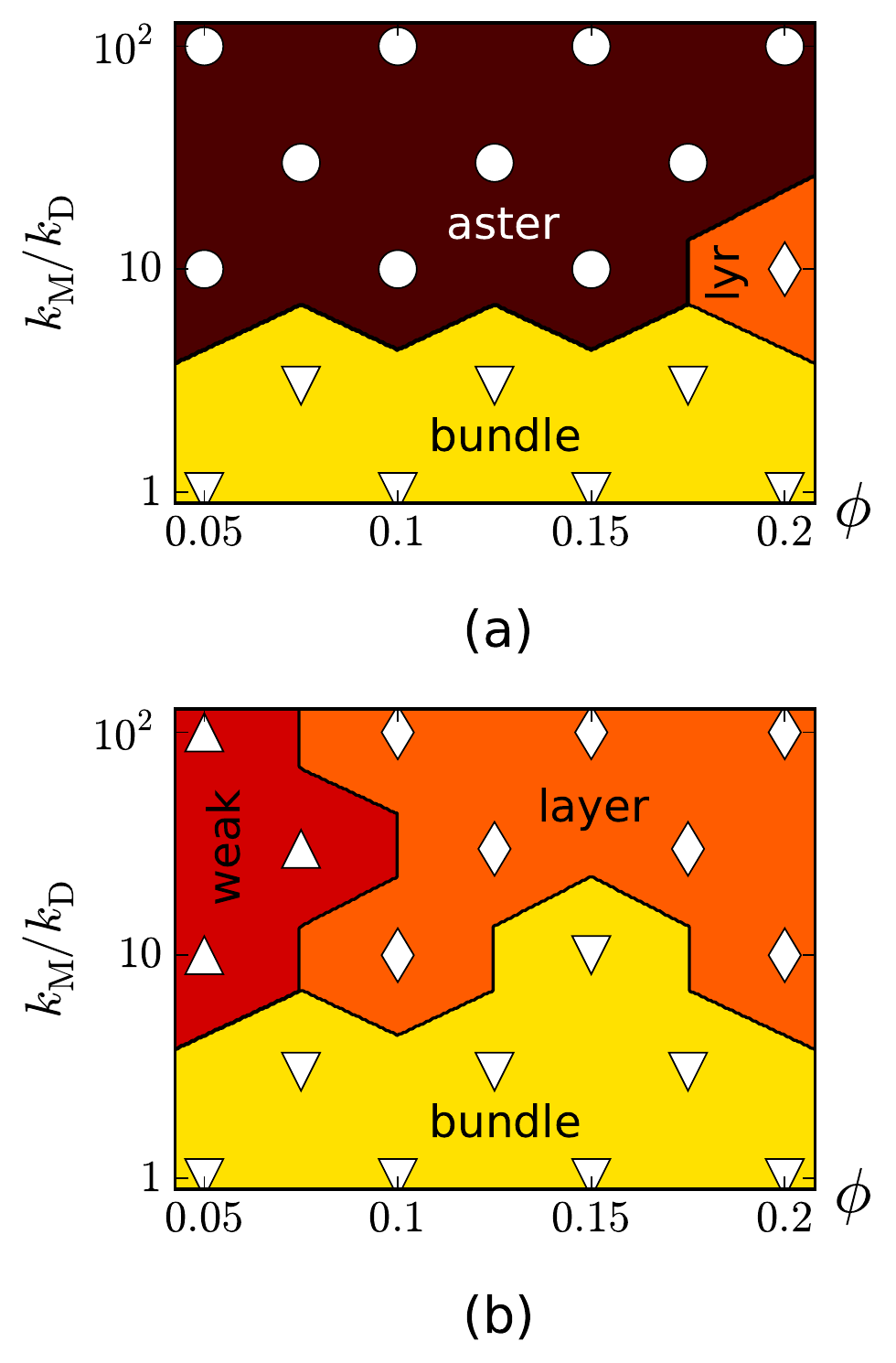}\hspace{1cm}}
\caption{States for filament density $\phi$ and motor speed $k_{\rm M}$ for (a)~$k_{\rm E}=k_{\rm D}$ and (b)~$k_{\rm E}=5k_{\rm D}$. $k_{\rm A}=40k_{\rm D}$ in both cases. Symbols refer to state: Circle (aster), diamond (layers), downward triangle (bundle) and upward triangle (weakly bound). The threshold parameters were $C^{\rm str}=0.9$ and~$\ell^{\rm str}=L/6$. Boundaries are drawn at midpoints between symbols.}
\label{f:state}
\end{figure}

The polarity correlation data can be used to define criteria to
determine the system state as follows: (i)~If $C_{pp}^{\perp}(r)$ remains above some threshold value $C^{\rm str}\approx1$ up to some given length $\ell^{\rm str}<L$, the state is regarded as strongly bound. (ii)~If a strongly bound state exhibits positive $C_{pp}^{\parallel}(r)$ and $C_{pp}^{\perp}(r)$ up to $r=L$, they are regarded as an aster or a layer; if not, they are a bundle. (iii)~Layers are differentiated from asters in that $C_{pp}^{\perp}$ remains non-negative up until the system size. Although clearly there is some arbitrariness in the choice of thresholds $C^{\rm str}$ and~$\ell^{\rm str}$, this only affects marginal cases near state boundaries. State diagrams for $k_{\rm A}=40k_{\rm D}$ are given in Fig.~\ref{f:state} for $k_{\rm E}=k_{\rm D}$ and $k_{\rm E}=5k_{\rm D}$.

\begin{figure}
\begin{center}
\includegraphics[width=8.5cm]{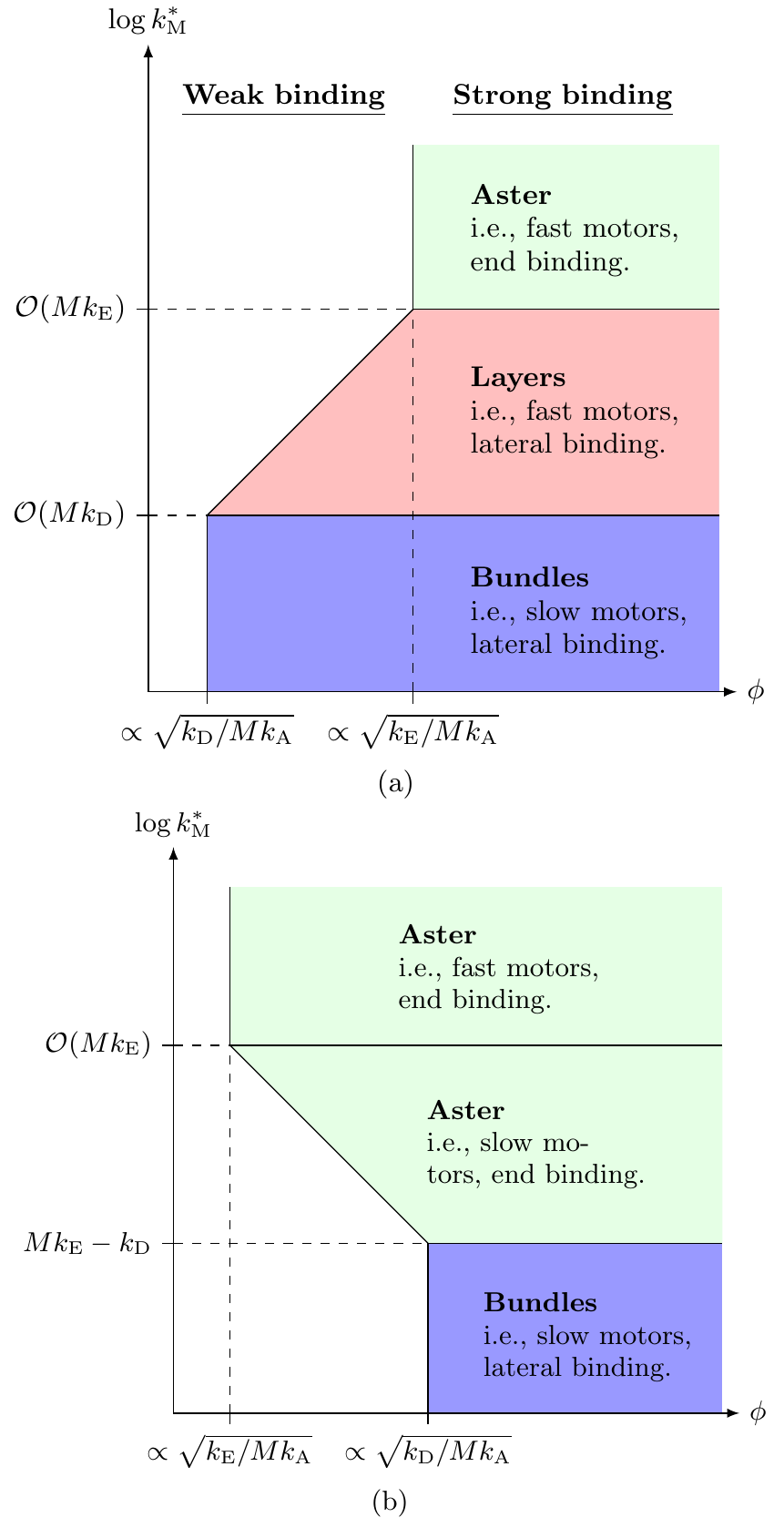}
\end{center}
\caption{(a) Schematic diagram denoting regimes predicted by the analytical model, here shown for $k_{\rm E}\gg k_{\rm D}$ (for $k_{\rm E}\approx k_{\rm D}$ the middle layers region vanishes). As described in the Appendix, the states for strong binding are predicted based on motor speed and the location of motor binding (end-dominated or laterally spread out). The boundary between weak and strong regimes is estimated by comparing the energies of thermal fluctuations and motor elasticity. To map to density, it has been assumed that $k_{\rm A}^{*}\propto k_{\rm A}\phi^{2}$.
(b) The same for $k_{\rm E}\ll k_{\rm D}$. Note that if $k_{\rm E}\leq k_{\rm D}/M$, the bundle region vanishes.
}
\label{f:thy_state}
\end{figure}

It is clear from Fig.~\ref{f:state} that reducing the dwell-time by increasing $k_{\rm E}$ favors layers over asters. To elucidate this crossover, we constructed and solved a one-filament model consisting of set of rate equations for the occupancy of motor heads along a filament, given known rates of motor attachment, detachment and movement. Since the actual attachment and movement rates depend on the current configuration, they are not known a priori, so to close the equations we assumed a constant attachment rate $k_{\rm A}^{*}$ and a constant movement rate $k_{\rm M}^{*}$. Details are given in the Appendix. Inspection of the solution reveals that the 
steady-state solution exhibits regimes for fast ($k_{\rm M}^{*}\gg Mk_{\rm D}$) and slow ($k_{\rm M}^{*}\ll Mk_{\rm D}$) motors, and also for end-dominated binding $2k^{*}_{\rm M}\gg k_{\rm E}M$ when most motors occupy $[+]$-ends. This latter regime corresponds to $t^{[+]}_{\rm occ}/t_{\rm occ}\gg M/2$. If we now assume that fast motors with end-dominated binding generate asters, fast motors without end-dominated binding generate layers ({\em i.e.}, $t^{[+]}_{\rm occ}/t_{\rm occ}\ll M/2$), and slow motors generate bundles, then the state diagram in Fig.~\ref{f:thy_state}(a) is predicted. Comparison to the numerical data in Fig.~\ref{f:state} reveals qualitative agreement, confirming the dominant factors determining pattern formation have been correctly identified. 
For $k_{\rm E}\ll k_{\rm D}$, lateral binding with fast motors is no longer possible, but end binding with slow motors can arise as shown in Fig.~\ref{f:thy_state}(b). This suggests the layers regime is replaced by an extended aster regime, consistent with the results of Fig.~\ref{f:stateMSDlowkA}.

\begin{figure}
\centerline{\includegraphics[width=8.5cm]{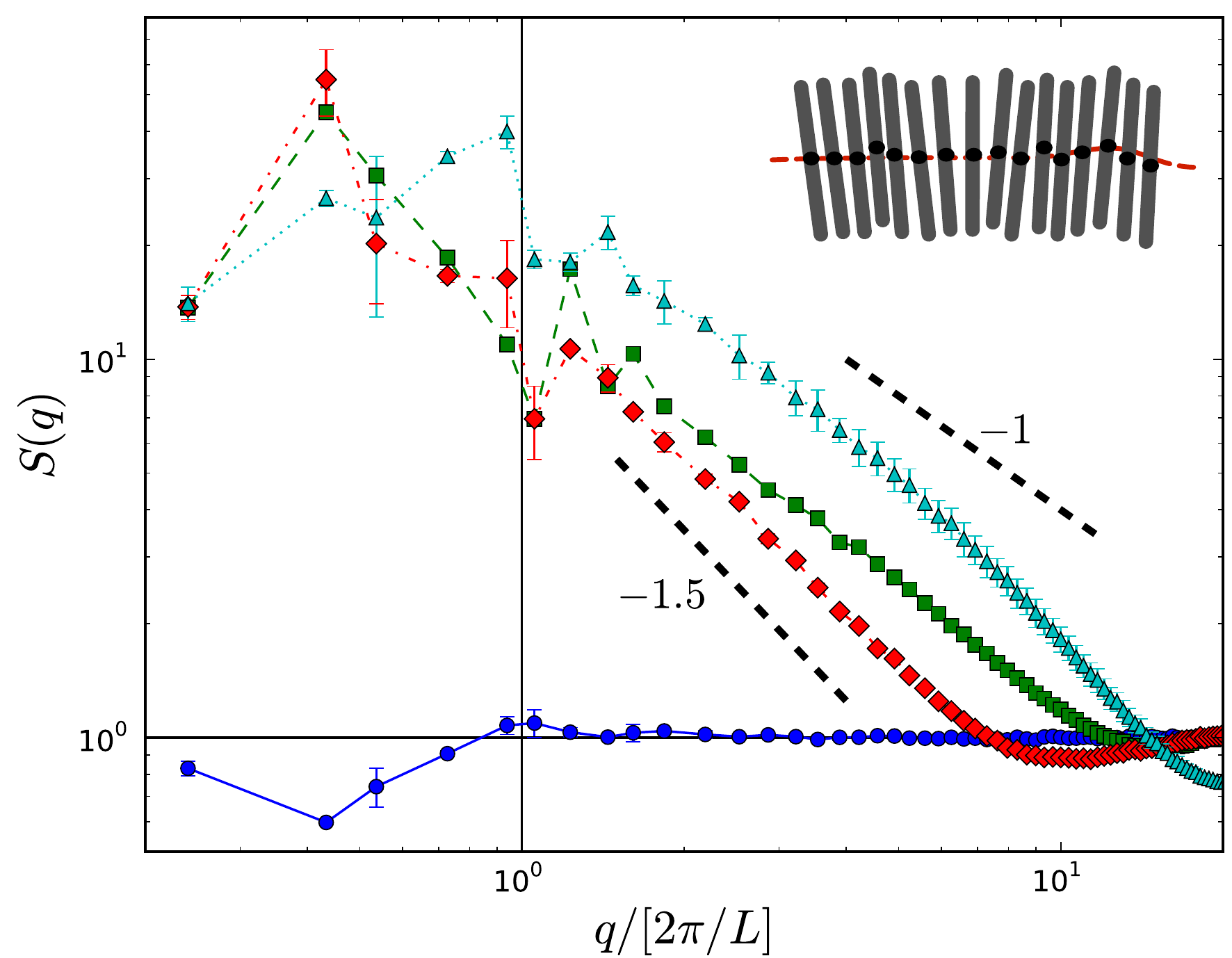}}
\caption{Static structure factor $S(q)$ for the same data (with the same symbols) as Fig.~\ref{f:polarCorrns}. The thick dashed lines have the given slope.
The schematic diagram in the inset explains why this $S(q)$ calculated from filament centers (black circles) can produce a similar spectrum to polymers.}
\label{f:Sq}
\end{figure}

For comparison to other active and passive systems, two further quantities often employed to characterize structural arrangements in disordered
or weakly ordered systems are now described. As shown in Fig.~\ref{f:Sq}, 
the static structure factor~$S(q)$, calculated from the 
correlations of filament centres, increases with decreasing wave vector $q$ 
for a broad range of~$q$. The variation is approximately a power law, $S(q)\propto q^{-\beta}$, with an exponent in the range $1\leq\beta<1.5$. Rod-like objects generate scattering curves with $\beta=1$~\cite{Roberts2012}; however, our 
structure factors $S(q)$ are calculated from the centres of mass of each 
filament and not the constituent monomers. Thus the power-law decay of $S(q)$ 
does not reflect the structure of a single filament, but rather arrays of laterally-aligned filaments as shown in the figure inset. Fluctuations in this array map to undulations in the line of centers, akin to a polymer in which each monomer corresponds to a filament's centre of mass, and indeed values of $\beta>1$ are expected for flexible polymers on lengths greater than their Kuhn length~\cite{EgalhaafWorkshop}. 

\begin{figure}
\centerline{\includegraphics[width=8.5cm]{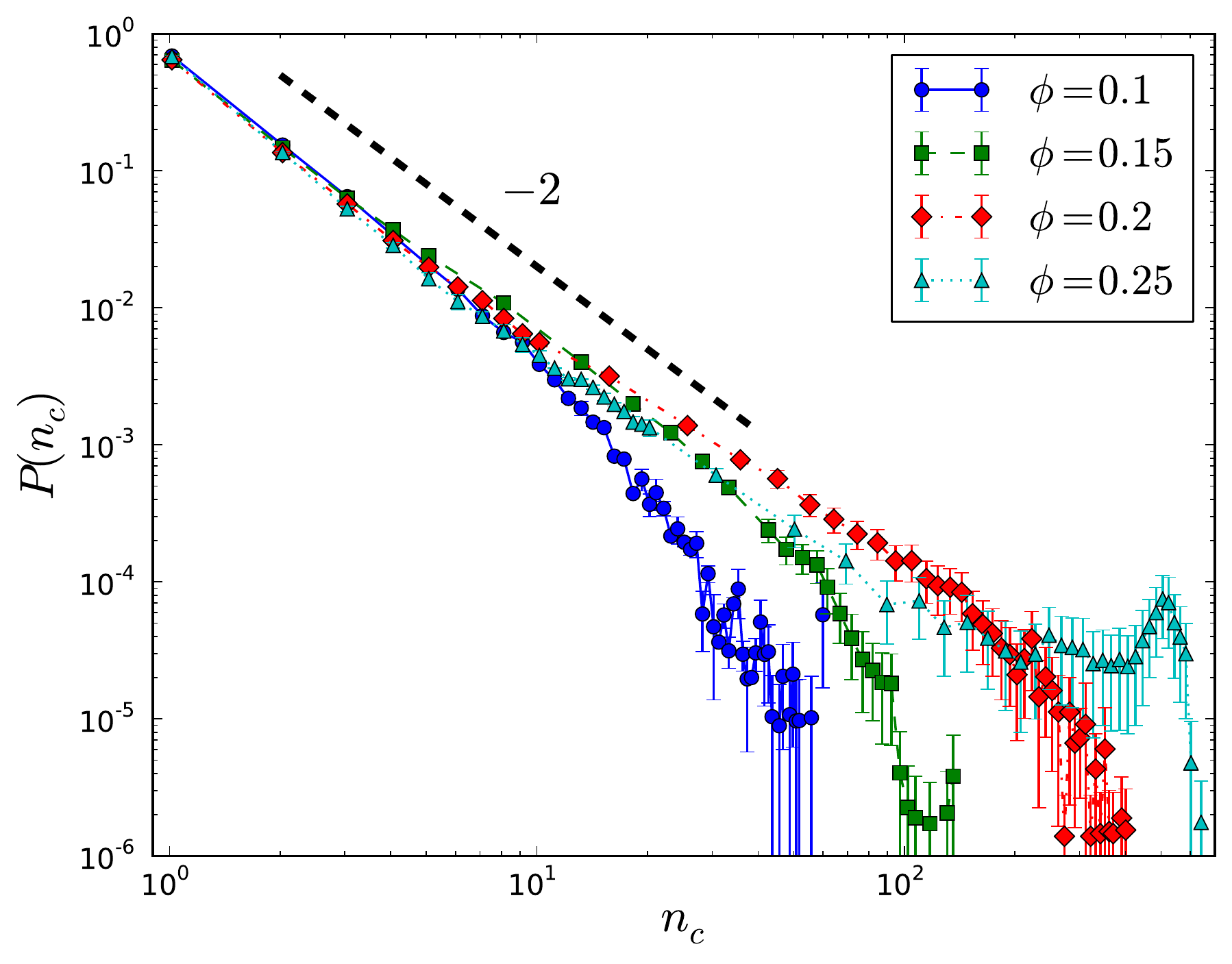}}
\caption{Probability density function $P(n_{c})$ of cluster sizes for $k_{\rm A}/k_{\rm D}=20$, $k_{\rm M}=10^{2}k_{\rm D}$, $k_{\rm E}=5k_{\rm D}$ and the filament densities $\phi$ given in the legend. The thick dashed line has a slope of~-2.}
\label{f:clusterDist}
\end{figure}

The weak and strong binding regimes are not distinct, and there is a continuous crossover between the two. This crossover  regime contains a scale-invariant distribution of cluster sizes $P(n_{\rm c})$, where two filaments are regarded as belonging to the same cluster if they are connected by at least one motor. As shown in Fig.~\ref{f:clusterDist}, $P(n_{c})$~is unimodal at small $n_{c}$ for weakly-bound states, becomes power law with an exponent $-2$ within the crossover, and bimodal for strongly-bound states. The exponent $-2$ is consistent with values observed for self-propelled particles in 2D~\cite{Yang2010,Chate2008}, but differs from the $-1$ observed in strictly 2D simulations of a similar model to here~\cite{Head2011b}.

\section{Discussion}
\label{s:discussion}

The use of microscopic modelling has highlighted the importance of a rarely considered microscopic parameter, namely the detachment rate from filament $[+]$-ends, in determining the motor-driven dynamics of weakly bound states, and the selection between asters and layers in the strongly bound regime. This parameter is not immediately accessible to ``hydrodynamic" theories. Furthermore, it 
cannot be easily varied experimentally, as it is an intrinsic property of 
motor proteins and filaments together, and is not amenable to 
continuous control (although see below). 
Microscopic modelling thus complements both ``hydrodynamic" theory and experiments by providing important 
insight that is difficult to gain by other means.

That the differential end-detachment rate $k_{\rm E}/k_{\rm D}$ can influence structure and dynamics implies that it may also modify the function of protein filament assemblies, and thus have been under the influence of natural selection, {\em i.e.}, a motor's $k_{\rm E}/k_{\rm D}$ may have evolved to increase the organism's fitness. If this speculation is true, it would suggest motor mutants exist with differing $k_{\rm E}/k_{\rm D}$, and creating such mutants in {\em in vitro} assays would help elucidate the role of dwell times in cellular function. 
It is also possible that other proteins binding to filament ends will affect the end-detachment rate.

Even if direct control over $k_{\rm E}/k_{\rm D}$ is not currently feasible, it should still be possible to test many of the predictions of our model using quasi-two-dimensional chambers, such as those that have been employed to study mixtures of microtubules and motors~\cite{Surrey2001,Sanchez2012}.
This geometry permits direct visualization of fluorescently-tagged filaments {\em via} light microscopy, allowing the quantities presented in Section~\ref{s:results} ({\em e.g.} the mean squared displacements in Figs.~\ref{f:msd_kA10}-\ref{f:msd_exps} and the polarity correlations in Fig.~\ref{f:polarCorrns}) to be extracted and compared to our predictions.
In addition, our predictions for scattering experiments are given in Fig.~\ref{f:Sq}.
However, experimental controls aligned with two of our key microscopic parameters, namely the ATP concentration (which modulates $k_{\rm M}$) and filament density, have not yet been systematically varied.
Surrey {\em et al.}~\cite{Surrey2001} only varied the motor concentration, related to our $k_{\rm A}$ (and indeed they found asters for high concentrations in agreement with our model), whereas Sanchez {\em et al.}~\cite{Sanchez2012} varied the ATP concentration but also added a depletion agent absent in our model.
We see no reason why these experiments could not be modified to directly test our predictions.
%

Hydrodynamic quantities defined on scales much longer than the filament length $L$ will require accelerated simulations before predictions can be made, as we now discuss.
In terms of the time for an unloaded motor to traverse a filament~$\tau_{L}=M/k_{\rm M}$, the total simulation times achieved varied from approximately $3\tau_{L}$ (for $k_{\rm M}=k_{\rm D}$) to $3\times10^{2}\tau_{L}$ (for $k_{\rm M}=10^{2}k_{\rm D}$). For actin-myosin systems $\tau_{L}\approx 0.1$s (based on $\approx1\mu$m filaments and motor speeds of $\approx 10\mu$m s$^{-1}$~\cite{HowardBook}), for which the maximum simulation time corresponds to minutes, shorter than typical experiments by 1-2 orders of magnitude. For kinesin-microtubule systems, $\tau_{L}\approx10$s ($\approx 10\mu$m filaments and motor speeds of $1\mu$m s$^{-1}$~\cite{HowardBook}), and here the simulation times approach hours, representative of experiments.

For length scales, however, the simulations fall short of the lengths orders of magnitude larger than $L$ required when coarse-graining ``hydrodynamic" 
equations~\cite{Liverpool2005}; all results presented here were for $X=Y\approx4L$. Experimental length scales are also typically much larger, except for cell-scale confinement where this model can already achieve comparable dimensions~\cite{Head2011b,Silva2011}. Roughly 90\% of our simulation time was spent performing the 
excluded-volume calculations (including Verlet list construction by cell sorting~\cite{AllenTildesly}), typically on 8-core shared memory architectures. This bottleneck can be reduced by extending the model to multi-node distributed architectures, or converting to run on many-core GPU devices. One order of magnitude improvement will allow box dimensions $X$, $Y\approx10L$ to be reached (with the same thickness $Z=L/6$), which would permit both length and time scales representative of {\em in vitro} microtubule-kinesin experiments to be replicated {\em in silico}.


In summary, simulations of a microscopic model of filament-motor 
mixtures qualitatively reproduce essential aspects of active gel properties. 
Improvements in simulation approaches will soon allow simulations on 
experimentally relevant time and length scales.

\begin{acknowledgments}
DAH was funded by a BHRC Senior Translational Research Fellowship, University of Leeds.
\end{acknowledgments}

%
%

%
%
\appendix*
\section{One-filament model}
\label{s:appendix}

It is possible to calculate the distribution of motor heads along a filament 
by introducing a model, in which the rates of attachment, 
detachment and motion are assumed to be constant in space and time. 
This is a simplification over the rules of Sec.~\ref{s:methods} for two 
reasons. Firstly, in the simulations the attachment rate depends on both 
the motor attachment rate $k_{\rm A}$ and the separation between 
monomers, and hence the local configuration of filaments. By assuming 
attachment occurs at a constant rate $k_{\rm A}^{*}$ which averages both 
factors, this coupling is neglected, leading to a significant simplification. 
Similarly the motor motion rate, which depends on the prefactor $k_{\rm M}$ 
and the change in motor elastic energy as per~Eq.~(\ref{e:move}), is 
reduced here to the constant value $k_{\rm M}^{*}$.
Coupled with the detachment rates $k_{\rm D}$ and $k_{\rm E}$, which are the same as in the simulations, these four rates allow the changes in occupation of motor heads along a filament to be fully determined.

The one-filament model is defined as follows. The rates for attachment $k_{\rm A}^{*}$, motion $k_{\rm M}^{*}$, and detachment $k_{\rm D}$ and $k_{\rm E}$ of motor heads are assumed to be constant and positive. As in Sec.~\ref{s:methods}, there is no excluded volume between motors.
Denoting the occupancy (mean number of motor heads) for each monomer by $n_{i}$, where $i=1$, $M$ corresponding to the $[-]$, $[+]$-ends respectively, then the rate equations are
\begin{eqnarray}
\partial_{t}n_{1}&=&k_{\rm A}^{*}-(k_{\rm M}^{*}+k_{\rm D})n_{1}\:,\nonumber\\
\partial_{t}n_{i}&=&k_{\rm A}^{*}+k_{\rm M}^{*} n_{i-1}-(k_{\rm M}^{*}+k_{\rm D})n_{i}\:,\quad1<i<M,\nonumber\\
\partial_{t}n_{M}&=&k_{\rm A}^{*}+k_{\rm M}^{*} n_{M-1}-k_{\rm E} n_{M}\:.
\label{e:master}
\end{eqnarray}
The steady-state solution $\partial_{t}n_{i}=0$ is
\begin{eqnarray}
n_{i}&=&\frac{k_{\rm A}^{*}}{k_{\rm D}}\left[1-\left(1+\frac{k_{\rm D}}{k_{\rm M}^{*}}\right)^{-i}\right]\:,\: 1\leq i<M,
\label{e:thy_ni}
\\
n_{M} 
&=&
\frac{k_{\rm A}^{*}}{k_{\rm E}}\left\{
1+\frac{k_{\rm M}^{*}}{k_{\rm D}}\left[
1-\left(1+\frac{k_{\rm D}}{k_{\rm M}^{*}}\right)^{-(M-1)}
\right]
\right\}\:,
\nonumber\\
\label{e:thy_nM}
\end{eqnarray}
which obeys $n_{i}>0$ $\forall i$. These $n_{i}$ also obey the net balance equation
\begin{equation}
Mk_{\rm A}^{*}=k_{\rm D} \sum_{i=1}^{M-1}n_{i}+k_{\rm E} n_{M}
\:.
\end{equation}
For later convenience, note that the total number of motors excluding those at the $[+]$-end is
\begin{equation}
\sum_{i=1}^{M-1}n_{i}
=
\frac{k_{\rm A}^{*}}{k_{\rm D}}\left\{
M-1+\frac{k_{\rm M}^{*}}{k_{\rm D}}\left[
\left(1+\frac{k_{\rm D}}{k_{\rm M}^{*}}\right)^{-(M-1)}-1
\right]
\right\}\\
\label{e:thy_sum_ni}
\end{equation}

Inspection of Eqs.~(\ref{e:thy_ni}), (\ref{e:thy_nM}) and (\ref{e:thy_sum_ni}) reveals different solution regimes for $k_{\rm M}^{*}\gg Mk_{\rm D}$ and $k_{\rm M}^{*}\ll Mk_{\rm D}$. We refer to these as the fast and slow motor regimes respectively. For fast motors, $Mk_{\rm D}/k_{\rm M}^{*}$ becomes a small parameter which can be expanded about, for which (\ref{e:thy_nM}) and (\ref{e:thy_sum_ni}) become (assuming $M\gg1$)
\begin{eqnarray}
\sum_{i=1}^{M-1}n_{i}
&\approx&
\frac{M^{2}k_{\rm A}^{*}}{2k_{\rm M}^{*}}\:,
\\
n_{M}
&\approx&
\frac{Mk_{\rm A}^{*}}{k_{\rm E}}\:.
\\\nonumber
\end{eqnarray}
Thus if $2k_{\rm M}^{*}\gg k_{\rm E} M$, $n_{M}\gg\sum_{i=1}^{M-1}n_{i}$ and almost all motors will be found at the $[+]$-end. This is referred to as end binding. Conversely, lateral binding arises when $2k_{\rm M}^{*}\ll k_{\rm E} M$ and most motors are at locations along the filament other than the $[+]$-end.
Note that this cannot happen if $k_{\rm E}\ll k_{\rm D}$.
Repeating this calculation for slow motors $k_{\rm M}^{*}\ll Mk_{\rm D}$ gives
\begin{eqnarray}
\sum_{i=1}^{M-1}n_{i}
&\approx&
\frac{Mk_{\rm A}^{*}}{k_{\rm D}}\:,
\\
n_{M}
&\approx&
\frac{k_{\rm A}^{*}}{k_{\rm E}}\left\{1+\frac{k_{\rm M}^{*}}{k_{\rm D}}\right\}\:.
\end{eqnarray}
Therefore slow motors produce lateral-dominated binding unless
\begin{equation}
\frac{k_{\rm E}}{k_{\rm D}}
\ll
\frac{1}{M}
+
\frac{k_{\rm M}^{*}}{Mk_{\rm D}}
\quad,
\end{equation}
when end-dominated binding arises. Note that the right hand side of this equation is much less than unity from the assumption of slow motors, $k_{\rm M}^{*}\ll Mk_{\rm D}$. Thus, end-binding with slow motors is only possible with reduced end detachment $k_{\rm E}\ll k_{\rm D}$.

In terms of occupancy times, it can be directly inferred from (\ref{e:master}) that $t^{[+]}_{\rm occ}/t_{\rm occ}=(k_{\rm D}+k_{\rm M}^{*})/k_{\rm E}$. For fast motors this simplifies to $t^{[+]}_{\rm occ}/t_{\rm occ}\approx k_{\rm M}^{*}/k_{\rm E}$, which corresponds to $t^{[+]}_{\rm occ}/t_{\rm occ}\gg M/2$ for end-binding, and $t^{[+]}_{\rm occ}/t_{\rm occ}\ll M/2$ for lateral binding. We note that these expressions shed no light on the empirical scaling variables $\x$ and $\y$ employed in Sec.~\ref{s:resDynamics}, and assume this analysis is too simplistic for dynamical quantities.

To influence filament organization, motors must first overcome the thermal motion of filaments. This can be estimated by comparing the total elastic energy of the motors to the thermal energy for filament motion. When the elastic energy dominates the thermal energy of filament motion, this is referred to as strong binding; the converse limit is weak binding. To estimate when each regime arises, note that the thermal energy of filament motion is of order $k_{\rm B}T$. For the elastic energy, using the same spring constant $k_{\rm B}T/b^{2}$ as in Sec.~\ref{s:methods} and assuming typical motor extensions of order~$b$, the total elastic energy is of order $k_{\rm B}T\sum_{i=1}^{M}n_{i}$. Strong binding is thus expected when $\sum_{i=1}^{M}n_{i}\gg 1$, which can be estimated for each regime discussed above using the corresponding expression for $\sum_{i=1}^{M}n_{i}$.


\end{document}